\documentclass[twocolumn,showpacs,preprintnumbers,amsmath,amssymb]{revtex4}
\usepackage{graphicx}% Include figure files
\begin{document}

\title{Cotunneling and one-dimensional localization in individual single-wall carbon nanotubes}

\author{B. Gao,$^1$ D.C. Glattli,$^{1,2}$ B. Pla\c{c}ais,$^1$ A. Bachtold$^{1,3*}$}

\address{
$^1$ Laboratoire Pierre Aigrain, Ecole Normale Sup\'{e}rieure, 24
rue Lhomond, 75231 Paris 05, France. $^2$ SPEC, CEA Saclay,
F-91191 Gif-sur-Yvette, France, $^3$ ICN and CNM-CSIC, Campus
UABarcelona, E-08193 Bellaterra, Spain.}

\begin{abstract}
We report on the temperature dependence of the intrinsic
resistance of long individual disordered single-wall carbon
nanotubes. The resistance grows dramatically as the temperature is
reduced, and the functional form is consistent with an activated
behavior. These results are described by Coulomb blockade along a
series of quantum dots. We occasionally observe a kink in the
activated behavior that reflects the change of the activation
energy as the temperature range is changed. This is attributed to
charge hopping events between non-adjacent quantum dots, which is
possible through cotunneling processes.

\end{abstract}

\vspace{.3cm} 

\date{ \today}
\maketitle

Single-wall carbon nanotubes (SWNT) are an excellent system to
study one-dimensional (1-d) transport. In particular, the effect
of disorder in 1-d is very pronounced; current lines have to
follow the wire and cannot go round impurity centers. As the
transmission of impurity centers becomes low enough, the 1-d wire
is divided in a series of quantum dots. The conduction is then
thermally activated $R(T) \approx \textmd{exp} (T^{-1})$
\cite{Ruzin,Staring,Chandrasekhar,Bezryadin}.

Measurements on 2-d or 3-d arrays of quantum dots can show a
slower than thermally activated dependence of the conduction $R(T)
\approx \textmd{exp} (T^{-0.5})$ \cite{Beverly,Yu}. This has been
recently attributed to cotunneling processes, which allow charge
transfer between non-adjacent quantum dots \cite{Tran,Feigelman,
Beloborodov}. Indeed, cotunneling transport in a series of quantum
dots is analogous to variable-range hopping (VRH)
\cite{Shklovskii}. Charges try to find hopping events with the
lowest activation energy and the shortest hopping distance. The
slower than thermally activated dependence of the conduction is
then a result of successive thermally activated curves with the
activation energy that decreases as the temperature is reduced.
However, such a succession of activated curves remain to be
observed.

Localization experiments have been carried on nanotube films or
individual SWNTs contacted to microfabricated electrodes, but
tube-tube junctions and tube-electrode interfaces make the
analysis difficult \cite{Bezryadin,Shea,Jang,Fuhrer,Vavro}. In our
experiments, the intrinsic resistance of disordered SWNTs is
measured in a four-point configuration \cite{Gao2}. The intrinsic
resistance is found to be thermally activated. As the gate voltage
($V_g$) is swept, we observe Coulomb blockade oscillations that
can be rather regular in some cases. These measurements are
consistent with a series of quantum dots that are typically
$\gtrsim 10$~nm long. Importantly, we also observe kinks in the
activated behavior of $R(T)$ that suggest the change of the
activation energy as the temperature range is varied. These kinks
are attributed to cotunneling processes.

\begin{figure}
\includegraphics{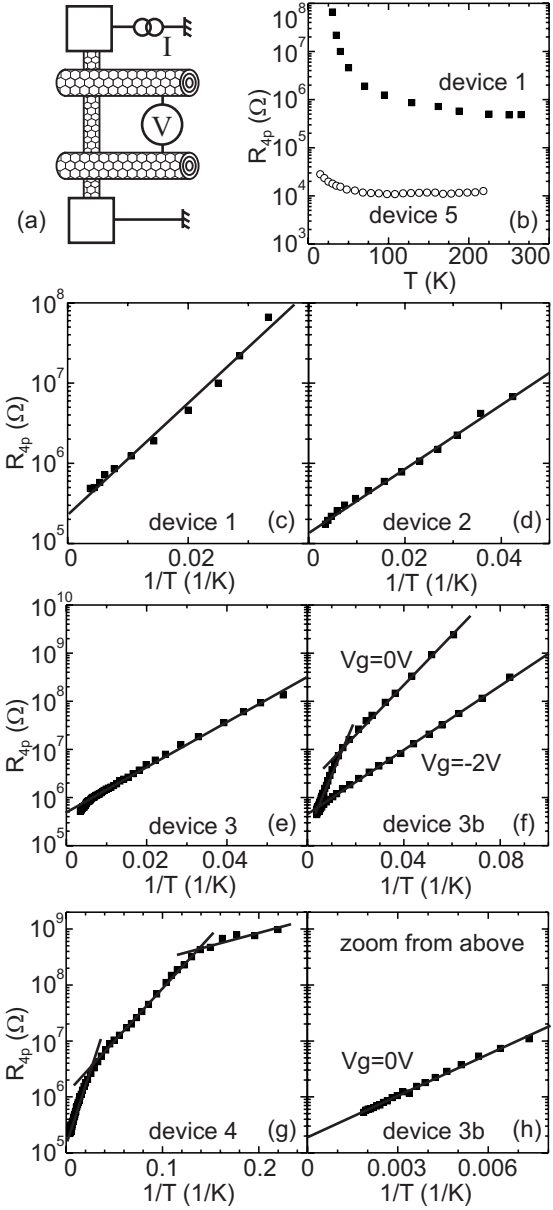}
 \caption{
Four-point resistance. (a) Device schematic. (b-h) $R_{4pt}$ as a
function of temperature. When the value of $V_g$ is not indicated,
$V_g=0$. Device 3b is the same device as Device 3, but it has been
measured one month before. Microscopic changes might have been
occurred in between. }
\end{figure}

The fabrication of SWNT devices for four-point measurements has
been described in Ref. \cite{Gao2}. Briefly, $\sim1$ nm diameter
SWNTs grown by laser-ablation \cite{Thess} are selected with an
atomic force microscopy (AFM). Noninvasive voltage electrodes are
defined by positioning two MWNTs above the SWNT using AFM
manipulation. Cr/Au electrodes are patterned for electric
connection using electron-beam lithography (Fig. 1(a)).
Characteristics of the devices are summarized in Table 1.

\begin{table}
\includegraphics{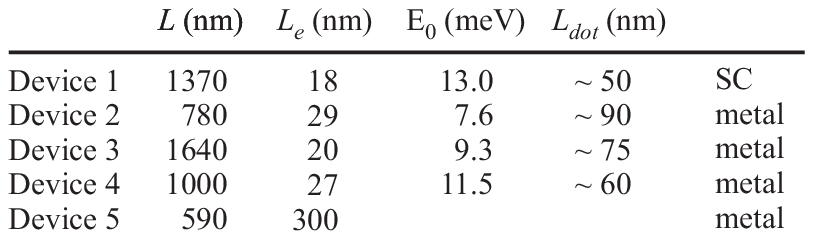}
 \caption{
Device characteristics. $L$ is the length between the MWNTs.
$L_{dot}$ is calculated from $E_0$ extracted at $V_g=0$. SC =
semiconducting tube with the threshold voltage at $\sim 40$ V.
 }
\end{table}

The four-point resistance $R_{4pt}$ of some SWNTs is particularly
large $> 100$ k$\Omega$ at 300~K. The nature of the scattering
centers responsible for this resistance is at present not
understood. Figure 1(b) shows the temperature dependence of
$R_{4pt}$ of one of those SWNTs (device 1). The curve is quite
flat at high $T$, while the resistance increases a lot below
100~K. The high-temperature resistance allows the estimation of
the elastic mean-free path $L_e$. Using $R_{4pt}=h/4e^2\cdot
L/L_e$ with $L$ the length between the MWNTs \cite{Gao2}, we get
$L_e=18$~nm.

For comparison, we also show a device that is significantly less
resistive at 300~K, $R_{4pt}=$12~k$\Omega$. The $R_{4pt}(T)$
variation is much less pronounced. This is consistent with
previously reported works on two-point, low ohmic SWNT devices,
where the $R_{2pt}(T)$ dependence is weak
\cite{Nygard,Liang,Kong}. For two-point devices with a large
resistance, the resistance has been reported to strongly grow as
$T$ goes to zero, which is usually associated to the change of the
contact resistance \cite{Bockrath}. In our case, the four-point
technique allows to separate the intrinsic and contact
resistances.

Figure 1(c) shows that the above measurement in the highly
diffusive tube is consistent with an activated behavior of the
resistance

%====================================================================
\begin{equation}
R_{4pt}=R_0\exp\frac{E_0}{kT}
\end{equation}
%====================================================================
with $E_0$ the activation energy. This dependency is observed in
other devices (see Fig. 1(d-f)). Similar $R(T)$ behaviors have
been reported for disordered wires microfabricated in
semiconductors \cite{Khavin,Kastner}.

Figures 1(f-h) show $R(T)$ measurements on other tubes. Some of
them deviate from the standard activated behavior. However, these
measurements can be described by successive exponential functions
with different activation energies, giving rise to kinks.
Interestingly, Fig. 1(f) shows that those two exponential
functions can merge in a single one on varying the gate voltage,
which is applied on the back side of the Si wafer. Overall, these
measurements suggest that the activation energy depends on the
temperature range and the gate voltage.

Further insight into transport properties is obtained by studying
the high-voltage regime. Figure 2 shows that the differential
$R_{4pt}$ is lowered as $V_{4pt}$ increases, and that the
dependence can be fitted with

%====================================================================
\begin{equation}
R_{4pt}=R_0\exp\frac{E_0-\alpha eV_{4pt}/2}{kT}
\end{equation}
%====================================================================

This suggests that an increase in the voltage reduces the
activation energy. An important point is that the slope deduced
from Fig. 2 gives $\alpha$ in Eq. (2) below unity. This means that
more than one energy barrier has to be overcome along the tube. A
rough estimate of the number of barriers $N$ can be made by taking
$N=1/\alpha$, which assumes identical barriers \cite{Khavin}. In
this way we obtain $N \lesssim 20$.

\begin{figure}
\includegraphics{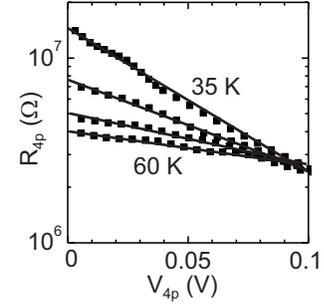}
 \caption{
Four-point differential resistance as a function of $V_{4pt}$ for
Device 1 at 35, 45, 55, and 60~K.}
\end{figure}

Fig. 3(a) shows the effect of the gate voltage, which controls the
position of the Fermi level in the tube. Large fluctuations of
$R_{4pt}(V_g)$ develop at low $T$ that look random
\cite{Kastner,Fowler}. At first sight, this may question the
activation behavior of $R_{4pt}(T)$ and the kinks discussed above.
However, Fig. 3(b) shows that a curve similar to Fig. 1(f)
($V_g=0$) is found by $V_g$ averaging $R_{4pt}(T)$. Moreover,
similar dependencies are observed, albeit with different
activation energies, for the minima and maxima of $R_{4pt}(V_g)$
as a function of $T$. This illustrates the robustness of the
activation behavior and the kink for a nanotube with a given Fermi
level.

\begin{figure}
\includegraphics{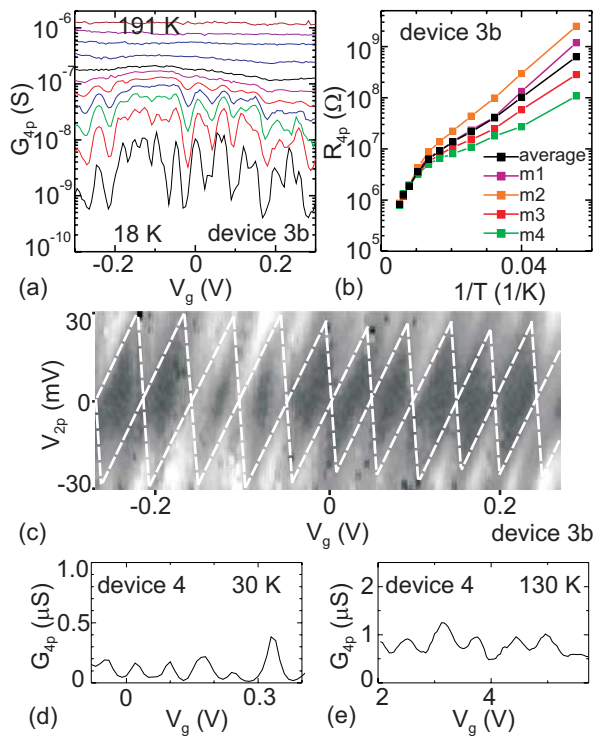}
 \caption{
Four-point resistance as a function of the gate voltage. (a)
$R_{4pt}(V_{g})$ for Device 3b at 18, 25, 31, 39, 49, 60, 74, 96,
125, 158, and 191~K. (b) $R_{4pt}(T)$ averaged over $V_g$ between
-0.3 and 0.3~V and taken at different conductance maxima
(m1@0.09V, m2@0.17V, m3@-0.25V, m4@-0.05V).  (c) Two-point
differential conductance as a function of $V_{g}$ and $V_{2pt}$ at
20~K. The same measurement with $G_{4pt}$ is very noisy.
$R_{4pt}(V_{g})$ and $R_{2pt}(V_{g})$ in the linear regime show
the same features at 20~K. Figures 1(f), 3(a) and 3(c) are taken
in three cooling runs. (d,e) $G_{4pt}(V_{g})$ for device 4. }
\end{figure}

While these fluctuations look random, oscillations can be found
that are quite regular within restricted $V_g$ ranges
\cite{Field}. Fig. 3(c) shows 10 successive oscillations. Note
that series of regular oscillations can be found at other $V_g$
ranges, and the period is then identical. Interestingly, Fig.
3(d,e) show that the period can change as the temperature is
modified. New oscillations can appear at lower $T$ that have a
shorter period.

We now discuss possible origins for the activated behavior of the
resistance. One possible mechanism is the Schottky barrier at the
interface between a metal electrode and a semiconducting nanotube
\cite{Martel}. However, we also observe the activated behavior in
metal tubes, which have no Schottky barriers. Moreover, the
four-point technique is aimed to avoid contributions from the
contacts \cite{Gao2}. Another mechanism is thus needed to account
for the results.

The fluctuations of $R_{4pt}(V_g)$ and the $R_{4pt}(T)$ dependence
may be attributed to universal conductance fluctuations and weak
localization. However, the variations of $R_{4pt}$ are much larger
than $h/e^2$, so that the results cannot simply originate from
interference corrections.

Strong localization (SL) is expected for highly diffusive systems
\cite{Shklovskii}. This theory has been used to explain
exponential length dependencies of the resistance measured in
nanotubes \cite{Pablo,Cumings,Gomez}. SL occurs when the
phase-coherence length $L_{\phi}$ becomes longer than the
localization length $L_{loc}$. This is equivalent to when the
width of the coherent states, $\hbar v_F/L_{\phi}$, becomes
smaller than the energy separation between the localized states.
The localized states are usually regarded as randomly distributed
in space and energy (see Fig. 4(a)) \cite{Imry,Fogler}. Irregular
oscillations of $R_{4pt}(V_g)$ are expected, which is in
opposition to our results.

\begin{figure}
\includegraphics{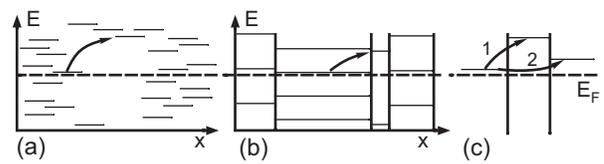}
 \caption{
Schematics of localized states along the nanotube. (a) States are
randomly distributed. (b) Strong barriers that define quantum
dots. (c) Proposed process to account for the kinks in Fig. 1
(f,g). Arrows represent hopping paths. }
\end{figure}

We now look at an alternative distribution of localized states as
schematized in Fig. 4(b). The tube is here divided in segments
separated by highly resistive scattering centers. The segment
lengths and therefore the energy separations can be different. At
high enough temperatures, levels are thermally smeared out except
for the shortest segment that has the largest level separations.
Oscillations of $R_{4pt}(V_g)$ are then regular, and the period is
large. At lower temperature, shorter periods arise from longer
segments, which agrees with experiments.

So far, the Coulomb interaction between electrons has not been
taken into account. However, the charging energy $E_c$ of a single
nanotube quantum dot is known to be larger than the level spacing
$\Delta E$ due to the geometrical confinement of the electron
wave. $\Delta E \approx 0.5 $ meV$\cdot \mu$m and the charging
energy for a tube dot connected to two tube leads is roughly $E_c
\approx 1.4 $ meV$\cdot \mu$m \cite{Bozovic}. This suggests that
the separation in energy between the localized states in Fig. 4(b)
is given by the charging energy.

Localization related to Coulomb blockade through multiple quantum
dots \cite{Ruzin,Staring,Chandrasekhar,Bezryadin} bears a lot in
common with the standard hopping model of the strong localization
theory \cite{Imry,Kastner,Fowler,Fogler}. Series of aperiodic
conductance oscillations are expected. Contrary to the SL regime,
however, quasi-periodic oscillations are also occasionally
predicted, in agreement with experiments. In addition the
resistance is expected to be thermally activated, which again
agrees with experiments. The activation energy is given by the dot
with the level that lies the furthest away from the Fermi level.
It may also be the largest separation of energy levels located in
neighboring dots. Thus, $E_0$ is expected to be gate voltage
dependent, consistent with our experimental findings.

We here estimate the size of the dots. The activation energy $E_0$
is roughly $0.5E_c$ of the shortest dot. $E_0$ is 11.5~meV for
device 4 at high $T$. Using $E_c \approx 1.4 $ meV$\cdot \mu$m, we
get a dot length of $\sim 60$~nm. Another possibility for this
estimation is to use the 625~meV period of the $R_{4pt}(V_g)$
oscillations at high $T$ (Fig. 3(e)). Indeed, $\Delta V_g\approx
12.5$~meV$\cdot \mu$m when looking Ref.\cite{Postma,Bozovic} for a
tube dot connected to two tube leads. This gives $\sim 20$~nm.
Note that $E_c$ cannot be estimated from the diamond height in
Fig. 3(c) since several dots lie in series. Finally, we obtain
$\sim 70$~nm by dividing the tube length by the dot number
obtained in Fig. 2. Those 3 estimations point all to quantum dot
lengths of a few 10~nm.

Table 1 gives the dot length of the other samples, estimated from
$E_0$. Dot lengths are slightly longer than the elastic length
$L_e$ determined at 300~K. $L_e$ corresponds to the separation
between scatterers when transmissions are 0.5. The barriers that
define the quantum dots thus have a transmission $\lesssim 0.5$,
which corresponds to a resistance $\gtrsim 6.5$~k$\Omega$. This is
consistent with the occurrence of Coulomb blockade since the
barrier resistance has to be larger than a few k$\Omega$s.

Having shown that the activated behavior of $R_{4pt}(T)$
originates from a series of quantum dots, we now turn our
attention to the kinks (Fig. 1(f-h)). This may simply come from
two thermally activated resistances that lie in series. However,
the activation energy would be higher at lower $T$, in opposition
to the measurements. Another mechanism is needed to describe the
kinks.

We propose that the kink is related to a mechanism that is
borrowed from the theory of variable range hopping
\cite{Shklovskii}, see Fig. 4(c). Electrons hop to the neighboring
quantum dot as indicated by the arrow 1. At lower $T$ it pays to
make the hop 2 to the second nearest quantum dot. The activation
energy is given by the level separation, which is thus reduced.
This is in agreement with the experiments.

In the VHR theory such hops are possible thanks to the tunneling
process. However, the tunnel probability is here dramatically low
since the second nearest dot is a few tens of nanometers far.
Another mechanism for the charge transfer between nonadjacent
quantum dots is needed to account for the results.

A possible mechanism is that the charge motion between two
nonadjacent dots occurs through cotunneling events
\cite{Tran,Feigelman, Beloborodov}. Cotunneling, which involves
the simultaneous tunneling of two or more electrons, transfers the
charge via a virtual state. This gives rise to current even when
the electron transport is Coulomb blockaded \cite{Pasquier}. A
cotunneling event is called inelastic when the quantum dot is left
in an excited state, and the event is otherwise called elastic.
For an individual quantum dot contacted to two leads, the
conductance contribution of elastic cotunneling is temperature
independent, while the contribution of inelastic cotunneling
scales as $T^2$.

Cotunneling in a series of quantum dots has been recently
calculated \cite{Feigelman, Beloborodov}. An energy reservoir
supplied by for e.g. phonons is required since $\epsilon _i$ the
energy of the initial state is most often different than $\epsilon
_f$ the energy of the final state (see hop 2 in Fig. 4(c)). The
resistance contribution between those two states is
\cite{Feigelman}

%====================================================================
\begin{equation}
R \propto R_0^N\exp\frac{\textmd{max}(|\epsilon _i-\epsilon
_f|,|\epsilon _i-\mu|,|\epsilon _f-\mu|)}{kT}
\end{equation}
%====================================================================
with $\mu$ the Fermi level and $N$ the number of dots between the
initial and the final states. $R_0=A_1E_c/(g\Delta E)$ for elastic
cotunneling and $R_0=A_2N^2E_c^2/(g(\epsilon _i-\epsilon _f)^2)$
for inelastic cotunneling with $g=Gh/e^2$ the average
dimensionless conductance of a barrier between two dots and $A_1$
and $A_2$ numerical constants of the order of unity. The coulomb
repulsion term between the dots $i$ and $f$ is here neglected for
simplicity. The prefactor $R_0^N$ grows as $N$ the number of
involved barriers gets larger. At high temperature, the hopping
process between two adjacent dots dominates transport and the
prefactor is low (hop 1 in Fig. 4(c)). As the temperature is
reduced, the exponential term grows a lot. It then pays to make
the hop between non-adjacent dots when the activation energy is
lower (hop 2 in Fig. 4(c)). This is consistent with the kinks
observed in Fig. 1.

The temperature $T^*$ of the first kink is expected to be around
$kT^* \simeq E_0^{above}-E_0^{below}$ with $E_0^{below}$ and
$E_0^{above}$ the activation energies below and above $T^*$. This
can be obtained from Eq. 3 taking into account that
$N^{below}-N^{above}=1$ and that $\ln R_0$ is of the order of
unity. This relation is consistent with the experiments. For
example, $E_0^{above}-E_0^{below}=14$ meV in Fig. 1(f) for $V_g=0$
while $kT^*=6$ meV.

We have seen that cotunneling processes allow a slower than
thermally activated dependence of the conduction. The main
contribution of the conduction comes from one (or a few) quantum
dot. The energy levels are randomly positioned in energy, so that
we cannot expect a specific functional form for the slower than
activated dependence measured here.

In conclusion, we have shown that the intrinsic resistance of
strongly disordered SWNTs is thermally activated. This is due to
Coulomb blockade in a series of $\gtrsim 10$~nm long quantum dots
lying along the tube. The activation energy is found to change as
the temperature range is changed. We attribute this result to
cotunneling processes. Disordered SWNTs form an interesting system
for future studies on one-dimensional localization. For example,
studies on longer tubes will be investigated to reach the 1-d
variable range hopping regime \cite{Fogler2}.

We thank C. Delalande for support, L. Forro for MWNTs, R. Smalley
for laser-ablation SWNTs, and P. McEuen, J.L. Pichard, M. Fogler
and M. Feigelman for discussions. LPA is CNRS-UMR8551 associated
to Paris 6 and 7. The research has been supported by ACN, Sesame.

$^{*}$ corresponding author: adrian.bachtold@cnm.es

\end{document}